\documentstyle[sprocl]{article}

\def\be{\beta}
\def\ga{\gamma}
\def\de{\delta}
\def\ep{\epsilon}

\def\ka{\kappa}
\def\la{\lambda}

\def\si{\sigma}

\def\ph{\phi}

\def\ch{\chi}
\def\ps{\psi}

\def\Ga{\Gamma}
\def\De{\Delta}

\def\cl{{\cal L}}

\def\prt{\partial}
\def\vev#1{\langle {#1}\rangle}

\def\fr#1#2{{{#1} \over {#2}}}
\def\frac#1#2{{\textstyle{{#1}\over {#2}}}}
\def\half{{\textstyle{1\over 2}}}
\def\lsim{\mathrel{\rlap{\lower4pt\hbox{\hskip1pt$\sim$}}
    \raise1pt\hbox{$<$}}}
\def\gsim{\mathrel{\rlap{\lower4pt\hbox{\hskip1pt$\sim$}}
    \raise1pt\hbox{$>$}}}
\def\sqr#1#2{{\vcenter{\vbox{\hrule height.#2pt
         \hbox{\vrule width.#2pt height#1pt \kern#1pt
         \vrule width.#2pt}
         \hrule height.#2pt}}}}

\def\lrprtmu{\stackrel{\leftrightarrow}{\partial_\mu}}
\def\lrprtnu{\stackrel{\leftrightarrow}{\partial^\nu}}

\def\Im{\hbox{Im}\,}

\newcommand{\beq}{\begin{equation}}
\newcommand{\eeq}{\end{equation}}
\newcommand{\bea}{\begin{eqnarray}}
\newcommand{\eea}{\end{eqnarray}}
\newcommand{\rf}[1]{(\ref{#1})}

\begin{document}

\begin{flushright}
IUHET 399\\
December 1998
\end{flushright}
\vglue 0.5 cm

\title{THEORY AND TESTS OF CPT AND LORENTZ VIOLATION\footnote
{Talk presented at CPT 98,
Bloomington, Indiana, November 1998}
}

\author{V.\ ALAN KOSTELECK\'Y}

\address{Physics Department, Indiana University\\
Bloomington, IN 47405, U.S.A.\\
Email: kostelec@indiana.edu} 

\maketitle\abstracts{ 
A general Lorentz-violating extension 
of the standard model of particle physics,
allowing for both CPT-even and CPT-odd effects,
is described.
Some of its theoretical aspects and experimental implications
are summarized.
}

\section{Introduction}

Nature appears to be covariant 
both under the discrete transformation CPT,
formed from the product of charge conjugation C,
parity inversion P, and time reversal T,
and under the continuous Lorentz transformations
including rotations and boosts.
The CPT theorem links these symmetries,
stating that under mild technical assumptions
CPT is an exact symmetry of local Lorentz-covariant field theories 
of point particles.\cite{cpt,sachs}

High-precision tests of both CPT and Lorentz invariance exist.
According to the Particle Data Group\cite{pdg}
the best figure of merit for CPT tests
involves the kaon particle-antiparticle mass difference,
which has been bounded by experiments at Fermilab and CERN
to\cite{kexpt}
\beq
\fr{|m_K - m_{\overline K}|}{m_K}
\lsim 10^{-18}
\quad .
\label{a}
\eeq
Indeed,
at present CPT is the only combination of C, P, T
observed as an exact symmetry of nature at the fundamental level.

The existence of high-precision experimental tests and
of the general CPT theorem for Lorentz-covariant particle theories 
means that the observation of CPT or Lorentz violation
would be a sensitive signal for unconventional physics
beyond the standard model.
It is therefore interesting to consider 
possible theoretical mechanisms 
through which CPT or Lorentz symmetry might be violated. 
Most suggestions along these lines in the literature
either have physical features
that seem unlikely to be realized in nature 
or involve radical revisions
of conventional quantum field theory, or both. 

Nonetheless,
there does exists at least one promising theoretical possibility,
based on spontaneous breaking of CPT and Lorentz symmetry
in an underlying theory,\cite{ks,kp}
that appears to be compatible both with experimental constraints
and with established quantum field theory.
It suggests that apparent breaking of CPT and Lorentz symmetry
might be observable in existing or feasible experiments, 
and it leads to a general phenomenology for 
CPT and Lorentz violation
at the level of the standard model
and quantum electrodynamics (QED).
The formulation and experimental implications
of this theory are briefly described in this talk.

\section{Framework}

In principle,
one can attempt to circumvent
the difficult issue of developing a
satisfactory microscopic theory allowing CPT and Lorentz breaking
by adopting a purely phenomenological approach.
This can be done by identifying 
and parametrizing observable quantities
that allow for CPT or Lorentz violation.

A well-known example is the phenomenology of CPT violation 
in oscillations of neutral kaons.\cite{leewu}
In the neutral-kaon system,
linear combinations of the strong-interaction eigenstates
$K^0$ and $\overline{K^0}$
form the physical eigenstates $K_S$ and $K_L$.
These combinations contain
two complex parameters, $\ep_K$ and $\de_K$,
parametrizing CP violation.
One, $\ep_K$, governs T violation with CPT symmetry 
while the other, $\de_K$, governs CPT violation with T symmetry.
The standard model of particle physics
has a mechanism for T violation,
and so $\ep_K$ is in this context 
nonzero and in principle calculable.
However, 
CPT is a symmetry of the standard model 
and so $\de_K$ is expected to vanish.
The possibility of a nonzero value of $\de_K$
is from this prespective only a phenomenological choice.
It has no grounds in a microscopic theory
and $\de_K$ is therefore not calculable.
Indeed,
in the absence of a microscopic theory,
it is even unclear whether this parametrization 
makes physical sense.
Moreover,
without a microscopic origin
$\de_K$ cannot be linked to other phenomenological parameters 
for CPT tests in different experiments.

Evidently,
it is more attractive theoretically
to develop an explicit microscopic theory 
for CPT and Lorentz violation.
With a theory of sufficient scope,
a general and quantitative  
phenomenology for CPT and Lorentz violation
could then be extracted 
at the level of the standard model.
This would allow 
calculation of phenomenological parameters,
direct comparisons between experiments,
and perhaps the prediction of signals.

The development of a microscopic theory of this type
is feasible within the context 
of spontaneous CPT and Lorentz breaking.\cite{ks,kp}
The idea is that the underlying theory of nature 
has a Lorentz- and CPT-covariant action,
but apparent violations of these symmetries 
could result from their spontanteous violation
in solutions to the theory.
It appears that this mechanism is viable from the theoretical viewpoint
and is an attractive way to 
violate CPT and Lorentz invariance.

Since spontaneous breaking is a property of the solution
rather than the dynamics of a theory,
the broken symmetry plays an important role in establishing the physics.
In the case of CPT and Lorentz violation,
spontaneous breaking has the advantage 
that many of the desirable properties of 
a Lorentz-covariant theory can be expected.
This is in sharp distinction to 
other types of CPT and Lorentz breaking,
which often are inconsistent 
with theoretical notions such as causality
or probability conservation.

The physics of a particle in a vacuum
with spontaneous Lorentz violation
is in some respects similar to that of
a conventional particle moving inside a biaxial crystal.\cite{cksm}
This system typically breaks Lorentz covariance 
both under rotations and under boosts.
However,
instead of leading to fundamental problems,
the lack of Lorentz covariance
is merely a result of the presence 
of the background crystal fields,
which leaves unaffected features such as causality.
Indeed,
one can explicitly confirm microcausality 
in certain simple models arising
from spontaneous CPT and Lorentz breaking.\cite{cksm}

In a Lorentz-covariant theory,
certain types of interaction among Lorentz-tensor fields
could trigger spontaneous breaking of Lorentz symmetry.
The idea is that these
interactions could destabilize the naive vacuum 
and generate nonzero Lorentz-tensor expectation values,
which fill the true vacuum 
and cause spontaneous Lorentz breaking.\cite{ks}
This also induces spontaneous CPT violation
whenever the expectation values
involve tensor fields with an odd number 
of spacetime indices.\cite{kp} 
Provided components of the expectation values lie 
along the four macroscopic spacetime dimensions,
apparent violations of CPT and Lorentz symmetry 
could arise at the level of the standard model.\cite{kp2}
This could lead to observable effects,
some of which are described in the following sections.

Conventional four-dimensional renormalizable gauge theories
such as the standard model
lack the necessary destabilizing interactions
to trigger spontaneous Lorentz violation.
However,
the mechanism may be realized in some string (M) theories
because suitable Lorentz-tensor interactions occur.
This can be investigated using string field theory
in the special case of the open bosonic string,
where the action and equations of motion
can be analytically derived 
for particle fields below some fixed level number $N$.
Obtaining and comparing solutions for different $N$
allows the identification of solutions 
that persist as $N$ increases.\cite{ks}
For some cases this procedure has been performed 
to a depth of over 20,000 terms in 
the static potential.\cite{kp}
The solutions remaining stable as $N$ increases
include ones spontaneously breaking Lorentz symmetry. 

In standard field theories,
spontaneous breaking of a continuous global symmetry
is accompanied by the appearance of massless modes,
ensured by the Nambu-Goldstone theorem.
Promoting a global spontaneously broken symmetry
to a local gauge symmetry leads to the Higgs mechanism:
the massless modes disappear
and a mass is generated for the gauge boson. 
Similarly,
spontaneous breaking of a continuous global Lorentz symmetry
would also lead to massless modes.
However,
although the inclusion of gravity
promotes Lorentz invariance to a local symmetry,
no analogue to the Higgs effect occurs.\cite{ks}
The dependence of the connection on
derivatives of the metric rather than the metric itself
ensures that the graviton propagator is affected
in such a way that no graviton mass is generated
when local Lorentz symmetry is spontaneously broken.

\section{Standard-Model and QED Extensions}

Assuming spontaneous CPT and Lorentz violation does occur,
then any apparent breaking 
at the level of the SU(3)$\times$SU(2)$\times$U(1) standard model
and QED must be highly suppressed 
to remain compatible with established experimental bounds.
If the appropriate dimensionless suppression factor 
is determined by the ratio of 
a low-energy (standard-model) scale
to the (Planck) scale of an underlying fundamental theory,
then relatively few observable effects 
of Lorentz or CPT violation would arise.
To study these,
it is useful to develop an extension of the standard model
obtained as the low-energy limit 
of the fundamental theory.\cite{kp2}

To gain insight about the construction of such an extension,
consider as an example
a possible coupling between one or more bosonic tensor fields
and fermion bilinears
in the low-energy limit of the underlying theory.
When the tensors acquire expectation values $\vev{T}$,
the low-energy theory gains additional terms of the form
\beq
\cl \sim \fr {\la} {M^k} 
\vev{T}\cdot\overline{\ps}\Ga(i\prt )^k\ch
+ {\textstyle h.c.}
\quad .
\label{aa}
\eeq
Here,
the gamma-matrix structure $\Ga$
and the $k$ spacetime derivatives $i\prt$
determine the Lorentz properties of the
bilinear in the fermion fields $\ps$, $\ch$
and hence fix the type of apparent CPT and Lorentz violation 
in the low-energy theory.
The effective coupling involves an expectation $\vev{T}$
together with a dimensionless coupling $\la$
and a suitable power of a large 
(Planck or compactification) scale $M$
associated with the fundamental theory.

Proceeding along these lines,
one can determine all possible terms arising 
at the level of the standard model
from spontaneous CPT and Lorentz breaking
in any underlying theory 
(not necessarily string theory).
This leads to
a general Lorentz-violating extension 
of the standard model
that includes both CPT-even and CPT-odd terms.\cite{cksm}
It contains 
all possible allowed hermitian terms preserving both
SU(3)$\times$SU(2)$\times$U(1) gauge invariance
and power-counting renormalizability.
It appears at present to be the sole candidate
for a consistent extension of the standard model
based on a microscopic theory of Lorentz violation.

Despite the apparent CPT and Lorentz breaking,
the standard-model extension exhibits several desirable properties of 
conventional Lorentz-covariant field theories
by virtue of its origin in spontaneous symmetry breaking
from a covariant underlying theory.\cite{cksm}
Thus, the usual quantization methods are valid
and features like microcausality and positivity of the energy
are to be expected. 
Also, 
energy and momentum are conserved provided 
the tensor expectation values are independent of spacetime position
(no soliton solutions).
Even one type of Lorentz symmetry remains: the theory is 
covariant under rotations or boosts of the observer's inertial frame
(observer Lorentz transformations).
The apparent Lorentz violations appear
only when (localized) fields are rotated or boosted
(particle Lorentz transformations)
relative to the vacuum tensor expectation values.

In the case of the conventional standard model,
one can obtain the usual versions of QED
by taking suitable limits.
For the standard-model extension,
it can be shown that the usual  
gauge symmetry breaking to the electromagnetic U(1) occurs,
and taking appropriate limits yields generalizations 
of the usual versions of QED. 
It turns out that the apparent CPT and Lorentz breaking 
can arise in both the photon and fermion sectors.\cite{cksm}
These extensions of QED are of particular interest
because many high-precision QED tests 
of CPT and Lorentz symmetry exist. 

An explicit and relatively simple example 
is the restriction of the standard-model extension
to an extension of QED involving 
only photons, electrons, and positrons.
The usual lagrangian is:
\beq
\cl^{\rm QED} =
\overline{\ps} \ga^\mu (\half i \lrprtmu - q A_\mu ) \ps 
- m \overline{\ps} \ps 
- \frac 1 4 F_{\mu\nu}F^{\mu\nu}
\quad .
\label{aaa}
\eeq
Apparent Lorentz violation
can occur in both the fermion and photon sectors,
and it can be CPT even or CPT odd.
The CPT-violating terms are:
\beq
\cl^{\rm CPT}_{e} =
- a_{\mu} \overline{\ps} \ga^{\mu} \ps 
- b_{\mu} \overline{\ps} \ga_5 \ga^{\mu} \ps \quad ,
$$
$$
\cl^{\rm CPT}_{\ga} =
\half (k_{AF})^\ka \ep_{\ka\la\mu\nu} A^\la F^{\mu\nu}
\quad .
\label{bbb}
\eeq
The CPT-preserving terms are:
\beq
\cl^{\rm Lorentz}_{e} = 
c_{\mu\nu} \overline{\ps} \ga^{\mu} 
(\half i \lrprtnu - q A^\nu ) \ps 
+ d_{\mu\nu} \overline{\ps} \ga_5 \ga^\mu 
(\half i \lrprtnu - q A^\nu ) \ps 
- \half H_{\mu\nu} \overline{\ps} \si^{\mu\nu} \ps 
$$
$$
\cl^{\rm Lorentz}_{\ga} =
-\frac 1 4 (k_F)_{\ka\la\mu\nu} F^{\ka\la}F^{\mu\nu}
\quad .
\label{ccc}
\eeq
The reader is referred to the literature\cite{cksm}
for details of the notation and conventions
and for information about the properties of the extra terms.
Note, however,
that all these terms are invariant under observer Lorentz
transformations,
whereas the expressions in Eqs.\ (4) and (5) violate
particle Lorentz invariance:
the coefficients of the extra terms 
behave as (minuscule) Lorentz- and CPT-violating couplings.
Note also that not all the components 
of the coefficients appearing are physically observable.
For example,
field redefinitions can be used to eliminate some 
coefficients of the type $a_\mu$
in the standard-model extension.
It turns out that these can be directly detected only 
in flavor-changing experiments,
and so they are unobservable at leading order 
in experiments restricted to electrons, positrons, and photons.

\section{Experimental Tests}

The standard-model extension described above 
forms a quantitative framework
within which various experimental tests of 
CPT and Lorentz symmetry
can be studied and compared.
Moreover,
potentially observable signals 
can be deduced in some cases.
Evidently,
any tests seeking to establish
nonzero CPT- and Lorentz-violating terms
in the standard-model extension
must contend with the expected heavy suppression 
of physical effects.

Although many tests of CPT and Lorentz symmetry 
lack the necessary sensitivity to possible signals,
a few special ones
can already place useful constraints on 
some of the new couplings in the standard-model extension.
Several of these tests are discussed elsewhere in
these proceedings.
Among the ones investigated to date are experiments with 
neutral-meson oscillations,\cite{kexpt,kp,kp2}$^{\!-\,}$\cite{ak}
comparative tests of QED 
in Penning traps,\cite{pennexpts,bkr}
spectroscopy of hydrogen and antihydrogen,\cite{antih,bkr2}
measurements of cosmological birefringence,\cite{cksm}
and observations of the baryon asymmetry.\cite{bckp}
The remainder of this talk provides a brief outline
of some of these studies. 
Other work is in progress,
including an investigation\cite{kla}
of constraints from clock-comparison experiments.\cite{cc}

\subsection{Neutral-Meson Oscillations}

Flavor oscillations occur or are anticipated
in a variety of neutral-meson systems,
including $K$, $D$, $B_d$, and $B_s$.
A neutral-meson state evolves in time
according to a non-hermitian two-by-two effective hamiltonian
in the meson-antimeson state space.
The effective hamiltonian involves complex parameters
$\ep_P$ and $\de_P$
that govern (indirect) CP violation,
where the neutral meson is denoted by $P$.
In the $K$ system,
$\ep_K$ and $\de_K$
are the same phenomenological quantities 
mentioned in section 2.
The parameter $\ep_P$ governs T violation,
while $\de_P$ governs CPT violation.
Bounds on CPT violation can be obtained 
by constraining the magnitude of $\de_P$ 
in experiments with meson oscillations.

In the context of the usual standard model,
which preserves CPT,
$\de_P$ is necessarily zero.
In contrast,
in the context of the standard-model extension
$\de_P$ is a derivable quantity.\cite{ak}
It turns out that
at leading order $\de_P$ depends only 
on a single type of extra coupling
in the standard-model extension.
This type of coupling has the form
$- a^q_{\mu} \overline{q} \ga^\mu q$,
where $q$ represents one of the valence quark fields
in the $P$ meson
and the quantity $a^q_{\mu}$ is spacetime constant 
but depends on the quark flavor $q$.

Since Lorentz symmetry is broken
in the standard-model extension,
the derived expression for $\de_P$
varies with the boost and orientation of the $P$ meson.\cite{ak}
Denoting by $\be^\mu \equiv \ga(1,\vec\be)$ 
the four-velocity of the $P$-meson 
in the frame in which the quantities $a^q_{\mu}$ are specified,
it can be shown that
$\de_P$ is given at leading order in all coupling coefficients 
in the standard-model extension by
\beq
\de_P \approx i \sin\hat\ph \exp(i\hat\ph) 
\ga(\De a_0 - \vec \be \cdot \De \vec a) /\De m
\quad .
\label{e}
\eeq
For simplicity,
subscripts $P$ have been omitted on the right-hand side.
In Eq.\ \rf{e},
$\De a_\mu \equiv a_\mu^{q_2} - a_\mu^{q_1}$,
where $q_1$ and $q_2$ are the valence-quark flavors 
for the $P$ meson.
Also,
$\hat\ph\equiv \tan^{-1}(2\De m/\De\ga)$,
where $\De m$ and $\De \ga$
are the mass and decay-rate differences,
respectively,
between the $P$-meson eigenstates.

This result has several implications.
One is that tests of CPT and Lorentz symmetry with neutral mesons
are independent at leading order of 
all other types of tests mentioned here.
This is because $\de_P$ is sensitive only 
to $a^q_{\mu}$
and because this sensitivity arises from flavor-changing effects.
None of the other experiments described here
involve flavor changes,
which can be shown to imply 
that none are sensitive to any $a^q_{\mu}$.

The result \rf{e} also makes predictions about signals 
in experiments with neutral mesons.
For example,
the real and imaginary parts of $\de_P$ 
are predicted to be proportional.\cite{kp2}
Similarly,
Eq.\ \rf{e} suggests
that the magnitude of $\de_P$ 
may be different for different $P$ 
due to the flavor dependence 
of the coefficients $a_\mu^q$.
For example,
if the coefficients $a_\mu^q$ grow with mass
as do the usual Yukawa couplings,
then the heavier neutral mesons
such as $D$ or $B_d$ may exhibit 
the largest CPT-violating effects.

The dependence of the result \rf{e}
on the meson boost magnitude and orientation
implies several notable effects
in the signals for CPT and Lorentz violation.\cite{ak}
For example,
two different experiments may have inequivalent 
CPT and Lorentz reach
despite having comparable statistical sensitivity.
This could arise if the mesons for one experiment
are well collimated while those for the other
have a $4\pi$ distribution,
or if the mesons involved in the two experiments have
very different momentum spectra.
Another interesting effect is the possibility of
diurnal variations in the data,
arising from the rotation of the Earth relative to 
the orientation of the coupling coefficients.\cite{ak}
This issue may be of some importance
because the data in neutral-meson experiments
are typically taken over many days.

At present,
the tightest clean experimental constraints on CPT violation
come from observations of the $K$ system.\cite{kexpt}
Some experimental results are now also available
for the heavier neutral-meson systems.
Two collaborations\cite{bexpt}
at CERN have performed analyses 
to investigate whether\cite{ckv}
existing data suffice to bound CPT violation.
The OPAL collaboration has published the measurement 
$\Im\de_{B_d} = -0.020 \pm 0.016 \pm 0.006$,
while the DELPHI collaboration has announced a preliminary 
result of $\Im\de_{B_d} = -0.011 \pm 0.017 \pm 0.005$.
Further theoretical and experimental studies are underway.

\subsection{QED Experiments}

High-precision measurements of properties of particles
and antiparticles can be obtained by
trapping individual particles for extended time periods.
Comparison of the results provides sensitive CPT tests.
Such experiments can constrain the couplings in the 
fermion sector of the QED extension.\cite{bkr}

Penning traps can be used to obtain
comparative measurements of particle and antiparticle 
anomaly and cyclotron frequencies.\cite{pennexpts}
The QED extension predicts direct signals and
also effects arising from diurnal variations
in the Earth-comoving laboratory frame.\cite{bkr}
Appropriate figures of merit for the various signals
have been defined
and the attainable experimental sensitivity estimated.

As one example,
comparing the anomalous magnetic moments
of electrons and positrons would generate an interesting bound
on the spatial components of the coefficient $b^e_\mu$
in the laboratory frame.
Available technology could place a limit of order $10^{-20}$ 
on the associated figure of merit.
A related test involves the search for diurnal variations
of the electron anomaly frequency,
for which a new experimental result
with a figure of merit bounded at $6\times 10^{-21}$
is presented in these proceedings
by Mittleman, Ioannou, and Dehmelt.\cite{rm} 
Analogous experiments with protons and antiprotons 
may be feasible.

Particle and antiparticle cyclotron frequencies can also be compared.
In these proceedings,
Gabrielse and coworkers 
present the results of
an experiment comparing the cyclotron frequencies of $H^-$ ions
and antiprotons in the same trap.\cite{gk}
The leading-order effects in this experiment
provide a test of Lorentz violation
in the context of the standard-model extension,
with an associated figure of merit 
bounded at $4\times 10^{-25}$.

Tests of CPT and Lorentz symmetry are also possible
via high-precision comparisons of spectroscopic data from trapped
hydrogen and antihydrogen.\cite{antih}
An investigation of the possible experimental signals
within the context of the standard-model and QED extensions
has been performed.\cite{bkr2}
Direct sensitivity to CPT- and Lorentz-violating couplings,
without suppression factors associated with the fine-structure constant,
arises for certain specific 1S-2S and hyperfine transitions 
in magnetically trapped hydrogen and antihydrogen.
In principle,
theoretically clean signals might be observed
for particular types of CPT and Lorentz violation.

The photon sector of the QED extension
can also be tightly constrained
from a combination of theoretical considerations
and terrestrial, astrophysical,
and cosmological experiments on electromagnetic radiation.
It is known that 
the pure-photon CPT-violating term in Eq.\ \rf{bbb} 
can generate negative contributions
to the energy,\cite{cfj}
which may limit its viability
and suggests the coefficient
$(k_{AF})^\ka$ should be zero.\cite{cksm}
In contrast,
the CPT-even term in the following equation
maintains a positive conserved energy.\cite{cksm}

The solutions of the extended Maxwell equations 
with CPT- and Lorentz-breaking effects
involve two independent propagating degrees of freedom,\cite{cksm}
as usual.
Unlike the conventional propagation of 
electromagnetic waves in vacuum,
however,
in the extended Maxwell case
the two modes have different dispersion relations.
This means the vacuum is birefringent.
Indeed,
the effects of the CPT and Lorentz violation 
on an electromagnetic wave traveling in the vacuum
are closely analogous to those exhibited by 
an electromagnetic wave in conventional electrodynamics 
that is passing through 
a transparent optically anisotropic and gyrotropic crystal 
with spatial dispersion of the axes.\cite{cksm}

The sharpest experimental limits on the extra coefficients
in the extended Maxwell equations
can be obtained by constraining the birefringence of radio waves 
on cosmological distance scales.
Considering first the
CPT-odd coefficient $(k_{AF})_\mu$,
one finds\cite{cfj,hpk}
a bound of the order of $\lsim 10^{-42}$ GeV
on its components.
A disputed claim exists\cite{nr,misc}
for a nonzero effect at the level of 
$|\vec k_{AF}|\sim 10^{-41}$ GeV.

For the CPT-even dimensionless coefficient $(k_F)_{\ka\la\mu\nu}$,
the single rotation-invariant irreducible component 
is constrained to $\lsim 10^{-23}$ 
by the existence of cosmic rays\cite{cg}
and other tests.
Rotation invariance is broken by all
the other irreducible components of $(k_F)_{\ka\la\mu\nu}$.
Although in principle it might be feasible to constrain
these coefficients with existing techniques for
measuring cosmological birefringence,\cite{cksm}
no limits presently exist.
It is plausible that 
a bound at the level of about $10^{-27}$
could be placed on components of $(k_F)_{\ka\la\mu\nu}$.

The sharp experimental constraints obtained on $(k_{AF})_\mu$ 
are compatible with the zero value 
needed to avoid negative-energy contributions.
However,
no symmetry protects a zero 
tree-level value of $(k_{AF})_\mu$.
It might therefore seem reasonable to expect 
$(k_{AF})_\mu$ to acquire a nonzero value 
from radiative corrections 
involving CPT-violating couplings
in the fermion sector.
Nonetheless,
this does \it not \rm occur:\cite{cksm}
an anomaly-cancellation mechanism can ensure that
the net sum of all one-loop radiative corrections is finite.
The situation is technically involved because 
the contribution from each individual radiative correction
is ambiguous,\cite{cksm,jk}
but the anomaly-cancellation mechanism can hold  
even if one chooses to define the theory
such that each individual radiative correction is nonzero.
Thus,
a tree-level CPT-odd term is unnecessary
for one-loop renormalizability.
Similar effects may occur at higher loops.
This ability to impose the vanishing 
of an otherwise allowed CPT-odd term 
represents a significant check 
on the consistency of the standard-model extension.

For the CPT-even Lorentz-violating pure-photon term
there is no similar mechanism,
and in fact calculations have explicitly demonstrated\cite{cksm}
the existence of divergent radiative corrections
at the one-loop level.
This therefore leaves open the interesting possibility of
future detection of a nonzero effect via measurements of
cosmological birefringence.

Various other possible observable CPT effects have been identified.
For example,
under suitable conditions
the observed baryon asymmetry can be generated in thermal equilibrium
through CPT- and Lorentz-violating bilinear terms.\cite{bckp}
A relatively large baryon asymmetry produced at grand-unified scales
would eventually become diluted to the observed value 
through sphaleron or other effects.
This mechanism represents one possible alternative 
to the conventional scenarios for baryogenesis,
in which nonequilibrium processes
and C- and CP-breaking interactions are required.\cite{ads}

\section*{Acknowledgments}
I thank Orfeu Bertolami, Robert Bluhm, Chuck Lane,
Don Colladay, Roman Jackiw, Rob Potting, Neil Russell, Stuart Samuel, 
and Rick Van Kooten for collaborations.
This work was supported in part
by the United States Department of Energy 
under grant number DE-FG02-91ER40661.

\end{document}